# Angular dependent ferromagnetic resonance of exfoliated yttrium iron garnet film under stress

Yufeng Wang, Peng Zhou[*], Shuai Liu, Yajun Qi[*], Tianjin Zhang[*]

Ministry of Education Key Laboratory for Green Preparation and Application of Functional Materials, Hubei Provincial Key Laboratory of Polymers, Collaborative Innovation Center for Advanced Organic Chemical Materials Co-constructed by the Province and Ministry, School of Materials Science and Engineering, Hubei University, Wuhan 430062, PR China

[*]E-mail addresses:

zhou@hubu.edu.cn (Peng Zhou),

yjqi@hubu.edu.cn (Yajun Qi)

zhangtj@hubu.edu.cn (Tianjin Zhang)

Yttrium iron garnet ($Y_3Fe_5O_{12}$, YIG) plays a significant role in the field of spintronics due to its low magnetic damping and insulating characteristics. However, most studies have focused on YIG in bulk form or as film grown on rigid substrates. In this study, YIG film has been exfoliated from two-layer-graphene covered (111) $Gd_3Ga_5O_{12}$ (GGG) substrate. Magnetic properties of YIG film under stress are investigated in detail via angular dependent ferromagnetic resonance. The relationship between magnetic parameters and compressive/tensile stress has been established. The findings of this work will be beneficial for the applications of flexible YIG film.

YIG is one of the most promising materials in the fields of spin insulatronics and magnon spintronics due to its low magnetic damping and long spin diffusion length[1-5]. Tuning magnetic parameters of YIG, including but not limited to magnetic anisotropy and spin wave, is the key to its applications in spin electronic devices.

The widely employed approach for tuning the magnetic properties of YIG films involves control and variation of thickness. Previous studies have demonstrated strong correlations of saturation induction, effective perpendicular magnetic anisotropy, damping constant, and inhomogeneity line broadening, with YIG film thickness[6]. Thicker films showed both smaller linewidths and smaller damping constants[7]. While spin Hall magnetoresistance and spin mixing conductance for thinner film are higher than that of thicker film[8]. However, film thickness variation is not an efficient way for film property tuning.

Introduction of stress/strain in YIG films is another method to tune their magnetic properties. The magnetoelectric coupling of YIG/piezoelectric bilayer composites could be significantly affected by the interfacial strain induced by converse piezoelectric effect[9-11]. Specifically, the electric-field-induced polarization switching and lattice strain in the $Pb(Mg_{1/3}Nb_{2/3})_{0.7}Ti_{0.3}O_3$ single crystal resulted in two different magnetization states, as well as manipulation of pure spin current transport in the YIG film[12]. However, it's difficult to directly grow high quality YIG film on single crystal piezoelectrics due to the structure and lattice differences. As magnetostrictive coefficient of YIG is relatively low[13], growing YIG film on flexible substrate (such as mica) is expected to realize even larger stress tuning magnetic properties than those grown on piezoelectric substrates [14]. While the epitaxial property of YIG film on mica still needs to be improved.

The results of this study include: (1) YIG film grown on (111) GGG substrate covered with bilayer graphene (denoted as YIG/2Gr/GGG) was exfoliated successfully. The high quality growth and exfoliation of YIG film were realized because of two reasons: atomic potential fields of GGG substrate could penetrate bilayer graphene, and the weak interaction between graphene layers enabled the exfoliation [15-18]. (2) The relationship between angular dependent ferromagnetic resonance (FMR) of exfoliated YIG and stress applied on YIG film was established. (3) Tensile and compressive stress exhibited significant impact on magnetic parameters, including gyromagnetic ratio ($\gamma$), effective magnetization ($4\pi M_{eff}$), surface perpendicular anisotropy field ($H_p$), and magnetic anisotropy constants ($K_{p1}$ and $K_{p2}$).

Prior to the film growth, the GGG substrate (length × width = 5 mm × 5 mm) surface was transferred bilayer graphene, which was carried out by Sixcarbon Tech Shenzhen. YIG film was grown by pulsed laser deposition (PLD) with a KrF laser source (wavelength of ≈ 248 nm). The ablation energy and repetition rate were 300 mJ and 5 Hz, respectively. The deposition was conducted under the following conditions: substrate temperature of 750 °C and oxygen partial pressure of 5 Pa. Subsequently, the YIG film was post-annealed at 800 °C for 3 h.

The phase and structure of YIG film were examined using X-ray diffraction (XRD, D8 Discover, Bruker, USA) with Cu Kα (λ = 1.5406 Å) as the radiation source. Compositional analysis was conducted using X-ray photoelectron spectroscopy (XPS, Escalab 250 Xi, Thermo Fisher, USA). Magnetic hysteresis loops were measured by Physical Property Measurement System (PPMS, Quantum Design DynaCool, USA). Ferromagnetic resonance (FMR) was characterized by electron spin resonance spectrometry (EMX Plus, Bruker, USA). All the measurements were carried out at room temperature.

Figure 1 shows the detailed fabrication process of YIG film growth and subsequent film exfoliation. To protect the bilayer graphene from oxidation, the YIG film was firstly grown without oxygen[15]. The proportion of time spent growing in an oxygen-free environment roughly constitutes 8.3% of the overall growth duration. A flexible PET with thickness of 0.1 mm was stuck on YIG film surface via epoxy, as shown in figure 1(c). Finally, YIG film stuck on PET was exfoliated from GGG substrate due to the weak interaction between graphene layers, as illustrated in figure 1(d).

Figure 2 shows the comparison of XPS and XRD spectra of YIG film before and after exfoliation, which indicates three points. (1) Both elements of Fe and Y, as well as YIG (444) diffraction peaks can be detected in YIG/2Gr/GGG and YIG/PET, which confirms successful exfoliation of YIG film from GGG substrate. (2) Diffraction peak of YIG (444) in figure 2(c) shifts to higher 2θ value after exfoliation, as shown in figure 2(f). Since the YIG film on PET was not as flat as that on GGG substrate during the XRD measurement, which inevitably caused the peak shift. One can also note that there are two more diffraction peaks located at 2θ of around 49° in figure 2(c). These two peaks could be (202) and (040) of $YFeO_3$[19-21], which was formed during the film deposition without oxygen. (3) For XPS spectra of Fe2p in figures 2(a) and (d), the peaks located at binding energy of ~710 eV and ~ 725 eV are attributed to Fe $2p_{3/2}$ and Fe$2p_{1/2}$, respectively[22-24]. There is a satellite peak at binding energy of ~719 in figure 2(a), indicating that the YIG film is mainly composed of Fe ion with valence of +3 [22]. However, the satellite peak in figure 2(d) is relatively weak, revealing the differences of valence of Fe ion in YIG. In fact, the XPS spectrum in figure 2(a) was collected on YIG surface, while the one in figure 2(d) was obtained from the opposite side of YIG (*i.e.* the interface between YIG and 2Gr/GGG before exfoliation). As YIG/2Gr/GGG was post-annealed at 800 °C for 3 h, Fe on YIG surface could be oxidized more thoroughly than the one on the opposite side. There might be higher content of $Fe^{2+}$ on the opposite side of YIG than on the surface. Therefore, deviations exhibit for binding energies of Fe2p and Y3d.

We firstly investigated the magnetic properties of YIG/PET without applying stress. Magnetic hysteresis loops shown in figure 3(a) indicate that easy axis is along film plane. Based on our previous measurements (which are not presented here), the thickness of YIG film is determined to be ~40 nm. Consequently, the saturation magnetization ($M_s$) is estimated to be 121±15 emu/cm$^3$, which is lower than the value of bulk YIG[25]. Two reasons could contribute to the low $M_s$ value. (1) YIG film was not peeled off from GGG substrate thoroughly due to the defects of graphene, which gave rise to lower magnetization. (2) YFeO$_3$ formed during oxygen-free deposition. Figure 3(b) shows in-plane angular dependent FMR spectra, where the bias magnetic field and microwave magnetic field are parallel and perpendicular to film plane, respectively. The direction of bias magnetic field does not significantly influence the in-plane FMR resonance fields. Dependence of out-of-plane FMR spectra on $\theta_H$ (the definition of $\theta_H$ is depicted in the inset of figure 3(d)) is shown in figure 3(c). For out-of-plane angular dependent FMR measurement, microwave magnetic field is parallel to film surface, while the angle between bias magnetic field and film normal varies. The corresponding FMR resonance field and linewidth are presented in figures 3(d) and (e), respectively. Three notes should be made about the out-of-plane FMR data. First, FMR field decreases with the increase of $\theta_H$, revealing that the easy axis is along film plane, in accordance with the observations made from magnetic hysteresis loops. Second, the FMR spectra exhibit a transition from a single peak to multiple peaks as $\theta_H$ deceases. Third, the FMR linewidth fluctuates within a range of 10 to 30 Oe.

As PET serves as a flexible substrate, it allows for the application of both compressive and tensile stresses to the YIG film when YIG/PET is bent, as illustrated in the inset of figures 4(a) and (d), respectively. We utilize curvature as a measure to quantify the extent of bending, with higher curvature values indicating larger stress being exerted on YIG film. In this work, the curvature values are 0.03, 0.07, and 0.1, with the asterisk (*) in the upper right corner indicating tensile stress. Since the in-plane FMR field does not change obviously with direction of magnetic field (figure 3(b)), only angular dependent of out-of-plane FMR spectra under different stresses were examined (as seen in figure S3). The corresponding FMR field and linewidth are exhibited in figures 4 and 5, respectively. The correlations between FMR field and $\theta_H$ at different curvature values are similar to result shown in figure 3(d).

As the Landau-Lifshitz-Gilbert equation and the free energy density of YIG film can be utilized to describe the magnetization dynamics, the relationship between FMR field and $\theta_H$ can be fitted using the following equation [6, 26-28].

$$\left(\frac{f}{\gamma}\right)^2 = [H\cos(\theta - \theta_H) - 4\pi M_{eff} \cos(2\theta) + H_p(3\sin^2\theta\cos^2\theta - \sin^4\theta)] \times [H\cos(\theta - \theta_H) - 4\pi M_{eff}\cos^2\theta + H_p\sin^2\theta\cos^2\theta] \quad (1)$$

Where $4\pi M_{eff} = 4\pi M_s - \frac{2K_{p1}}{M_s}$, $H_p = \frac{4K_{p2}}{M_s}$. $f$ is the FMR frequency (9.8 GHz in this work). $H$ is the bias magnetic field of FMR measurement. The angles $\theta$ and $\theta_H$ represent the orientation of equilibrium magnetization and the direction of $H$ with respect to the film normal, respectively.

The angular dependence of FMR field is fitted well using equation (1), as shown in figures 3(d) and 4. The magnetic parameters, $\gamma$, $4\pi M_{eff}$, $H_p$, $K_{p1}$ and $K_{p2}$, are obtained from the fitting, as illustrated in figure 6. To assess the changes in magnetic properties of YIG film before and after exfoliation, FMR spectra of YIG/2Gr/GGG were also collected and analyzed, as seen in figures S1 and S2. Two main results are evident from the FMR data.

First, YIG film does not exhibit obvious in-plane magnetic anisotropy, as evident in figures S1(a) and 3(b). Generally, epitaxial YIG film exists a three-fold magneto-crystalline anisotropy for (111) orientation, as well as a two-fold anisotropy arises due to the angle formed between plasma plume axis and substrate normal during PLD deposition[7]. This could imply the simultaneous presence of both three-fold magneto-crystalline anisotropy and two-fold anisotropy in the YIG film. Moreover, the existence of $YFeO_3$ could also contribute to weak in-plane anisotropy. On the other hand, multiple FMR peaks appear only at $\theta_H = 20°$ and $30°$ for YIG/2Gr/GGG, as illustrated in figure S1(b). In contrast, the exfoliated YIG film exhibits multiple peaks at most of the $\theta_H$ values, as seen in figures 3(a) and S3 (The symbols in figures S2(b) and 5 allow for the identification of single and multiple FMR peaks). In addition, the FMR linewidth of YIG/2Gr/GGG shows an initial decrease followed by an increase with the increment of $\theta_H$, while random FMR linewidths are observed with the variation of $\theta_H$ in exfoliated YIG film under stresses.

The emergence of multiple FMR peaks can be understood based on the volume inhomogeneity model[29-31]. Typically, the resonance frequency of the magnetic moment precession at the film surface is different from that of its bulk, which is due to the exchange coupling of spins occurring specifically at the film surface. At particular critical $\theta_H$ value, the uniform precession resonance of the film surface is identical to that of the film bulk, resulting in the detection of a single FMR peak[29]. The critical $\theta_H$ values of YIG/2Gr/GGG and YIG/PET (without applying stress) are 40° and 70°, respectively. Under compressive stress, the critical $\theta_H$ values are 90°, 60° and 60° when the curvature values are 0.03, 0.07, and 0.1, respectively. Interestingly, under tensile stress, there is no critical $\theta_H$ value for the curvature of 0.1*, as merely multiple FMR peaks are observed. While the critical $\theta_H$ values are 60° and 70° when the curvature values are 0.03* and 0.07*, respectively. The variation of critical $\theta_H$ values with the change of curvatures implies that the exfoliation and applying stress have a notable impact on the precession resonance of the film surface.

In addition, three contributions should be considered for FMR linewidth: (1) intrinsic linewidth due to Gilbert damping, (2) extrinsic contribution due to two-magnon scattering, (3) inhomogeneous linewidth broadening due to the inhomogeneity of the film[26]. Except for the case at curvature of 0.1*, the FMR linewidth at $\theta_H = 0°$ is larger than that at $\theta_H = 90°$, as observed in figures S2(b), S3(e), and 5. Given that two-magnon scattering is not allowed in magnetic thin film at $\theta_H = 0°$ ($H$ is perpendicular to film normal), in can be inferred that the impact of two-magnon scattering on FMR linewidth of YIG film is insignificant[6]. While the YIG film exhibits a contribution from inhomogeneous linewidth broadening, evidenced by the larger FMR linewidth at $\theta_H = 0°$ compared to $\theta_H = 90°$.

Second, $\gamma$ increases from 2.5 GHz/kOe (corresponding Landé g factor is 1.76) to 2.6 GHz/kOe (g = 1.86) as the increment of compressive stress, whereas it diminishes from 2.6 GHz/kOe to 2.4 GHz/kOe (g =1.71) as the increment of tensile stress, as depicted in figure 6(a). Specifically, the $\gamma$ values are 2.8 GHz/kOe (g =2.00) and 2.6 GHz/kOe for YIG/2Gr/GGG and unbent YIG/GGG, respectively. There is a correlation between g factor and the ratio of orbital and spin moments ($\mu_L/\mu_S$), where $\mu_L/\mu_S = (g-2)/2$ [26, 32]. Therefore, the exfoliation process leads to a reduction in the spin-orbit coupling of YIG film, while an increase in compressive stress results in an enhancement of the spin-orbit coupling.

As both compressive and tensile stresses increase in YIG/PET, a decrease is observed in $4\pi M_{eff}$, as shown in figure 6(b). The $4\pi M_{eff}$ value for YIG/2Gr/GGG is approximately 1162 Oe, which is much lower than the previously reported value [33]. This relatively low $4\pi M_{eff}$ value may be attributed to the oxygen-free PLD deposition, as well as the interfacial strain induced by GGG substrate. Tensile stress has a more prominent effect on $H_p$ than compressive stress, as seen in figure 6(c). When the curvature shifts from 0.03* to 0.1*, the change in $H_p$ reaches 1710 Oe. Negative $H_p$ values reveal that the easy axis is in the YIG film plane.

With the increase of tensile stress, $K_{p1}$ increases from -3.5×10$^4$ erg/cm$^3$ (unbent) to 3.6×10$^4$ erg/cm$^3$ (curvature of 0.1*), while $K_{p2}$ decreases from -3.1×10$^4$ erg/cm$^3$ (unbent) to -8.9×10$^4$ erg/cm$^3$ (curvature of 0.1*), as illustrated in figures 6(d) and (e). In addition, as compressive stress increases, $K_{p1}$ exhibits an upward trend, while the relationship between $K_{p2}$ and compressive is relatively weak.

Three points need to be noted. (1) In some FMR spectra, particularly when the YIG film is under stress, multiple FMR peaks are observed. The analyses and fitting are confined solely to the narrow central segment of the primary resonance. (2) FMR frequency in this work is kept at 9.8 GHz, further FMR measurement at various frequency could be carried out to investigate the magnetic damping of YIG film under stress. (3) Except for bending, stretching and compressing can also apply tensile and compressive stresses, respectively, resulting in varied effects on the magnetic properties of YIG film.

In summary, YIG film was exfoliated from two-layer-graphene covered GGG substrate successfully. The relationship between magnetic properties of YIG/PET and stress has been investigated via angular dependent FMR. The exfoliated YIG film exhibits in-plane magnetic anisotropy. Compressive and tensile stresses have significant impact on magnetic parameters, including γ, $4\pi M_{eff}$ and $H_p$. This work paves the way for the research of spintronic devices based on flexible YIG thin films.

Acknowledgements:

This research was funded by the National Natural Science Foundation of China (No. 12104137, No. 12374083, and No. 12474083), the China Postdoctoral Science Foundation (No. 2020M672315), the Program of Introducing Talents of Discipline to Universities ("111 Project", D18025) China, and the Program of Hubei Key Laboratory of Ferro- & Piezoelectric Materials and Devices (No. K202013).

Figure captions:

Figure 1 Fabrication process of exfoliated YIG film

Figure 2 XRD and XPS spectra of YIG film before (top panel, YIG film was on GGG) and after (bottom panel, YIG film was on PET) exfoliation. (a) and (d) XPS spectra of Fe2p, (b) and (e) XPS spectra of Y3d, (c) and (f) XRD spectra of YIG/2Gr/GGG and YIG/PET

Figure 3 (a) magnetic hysteresis loops and (b) angular dependent of in-plane FMR spectra of the exfoliated YIG film without applying stress. (c) Out-of-plane FMR spectra of the exfoliated YIG film without applying stress, the angle of $\theta_H$ was varied in an increment of 10°. Dependence of FMR field (d) and FMR linewidth (e) on $\theta_H$, the data in (d) is obtained from (c) Inset in (a) is diagram of YIG/PET. Angles in (b) are relative to an edge of YIG film. The red line in (d) is a fit to equation (1)

Figure 4 FMR field as a function of $\theta_H$ for YIG/PET at curvature of (a) 0.03, (b) 0.07, (c) 0.1, (d) 0.03*, (e) 0.07*, (f) 0.1*

Figure 5 FMR linewidth as a function of $\theta_H$ for YIG/PET at curvature of (a) 0.03, (b) 0.07, (c) 0.1, (d) 0.03*, (e) 0.07*, (f) 0.1*

Figure 6 Magnetic parameter as a function of curvature. (a) γ, (b) $4\pi M_{\text{eff}}$, (c) $H_p$, (d) $K_{p1}$, and (e) $K_{p2}$

Figure 1

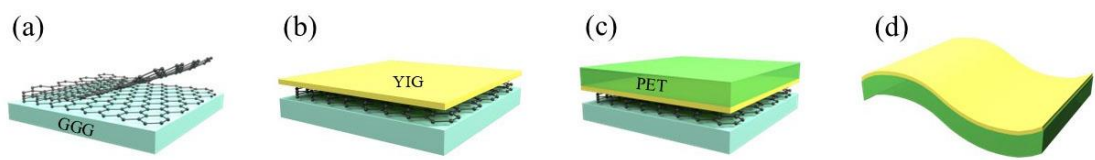

Figure 2

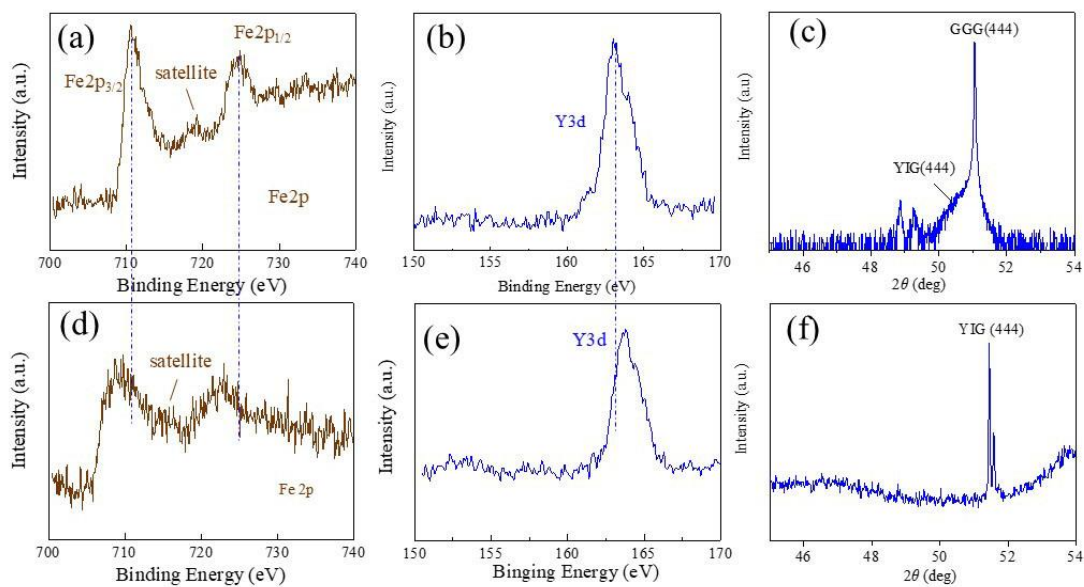

Figure 3

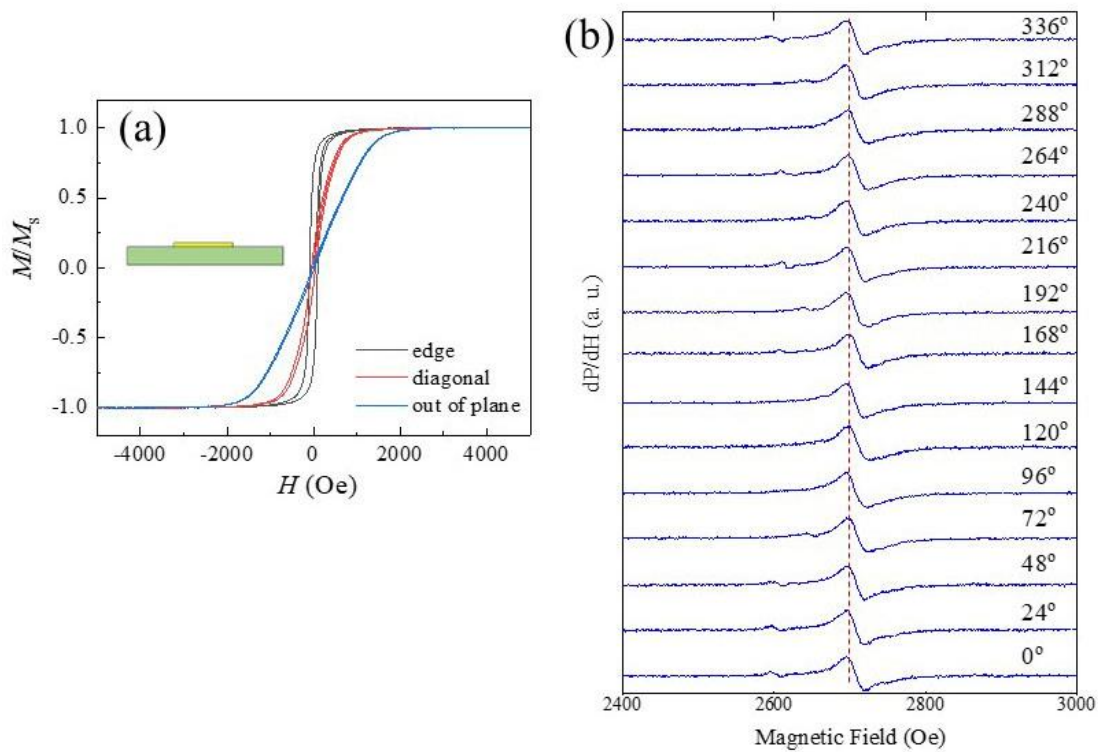

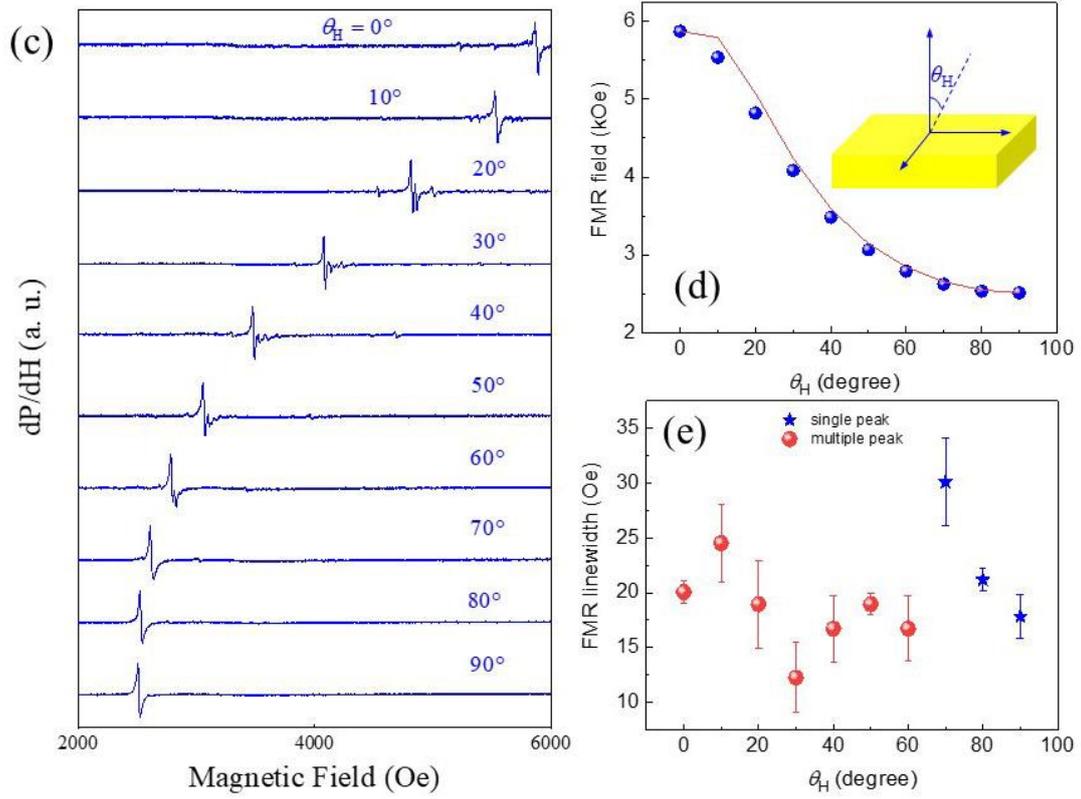

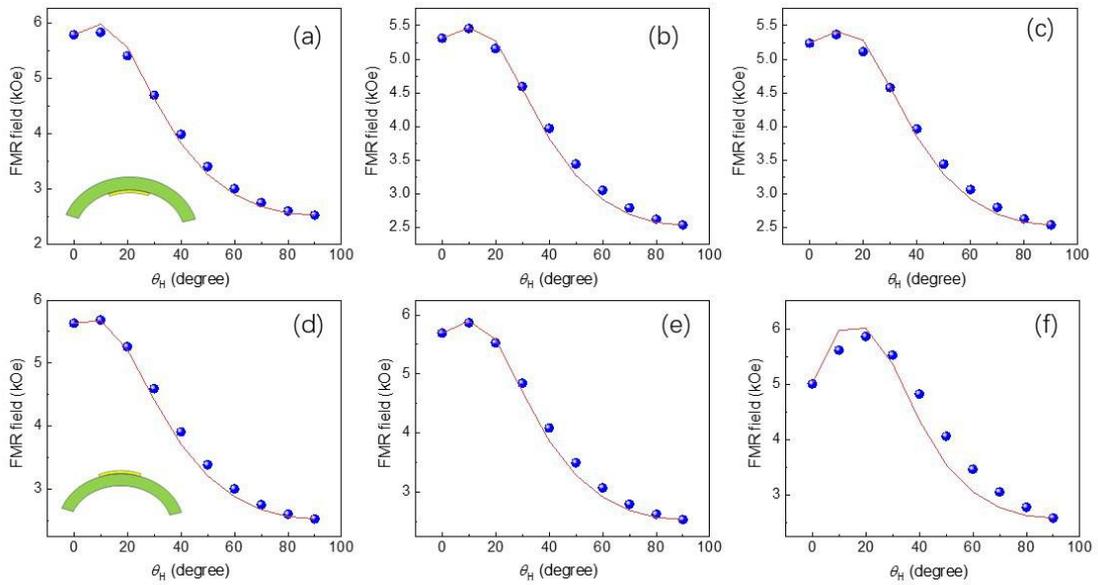

Figure 4

Figure 5

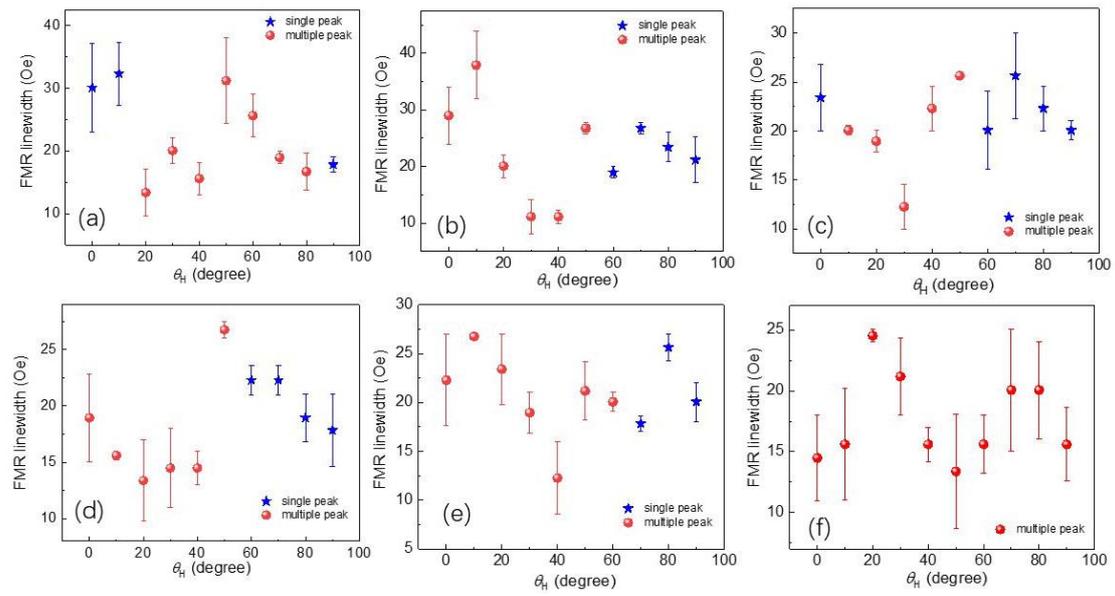

Figure 6

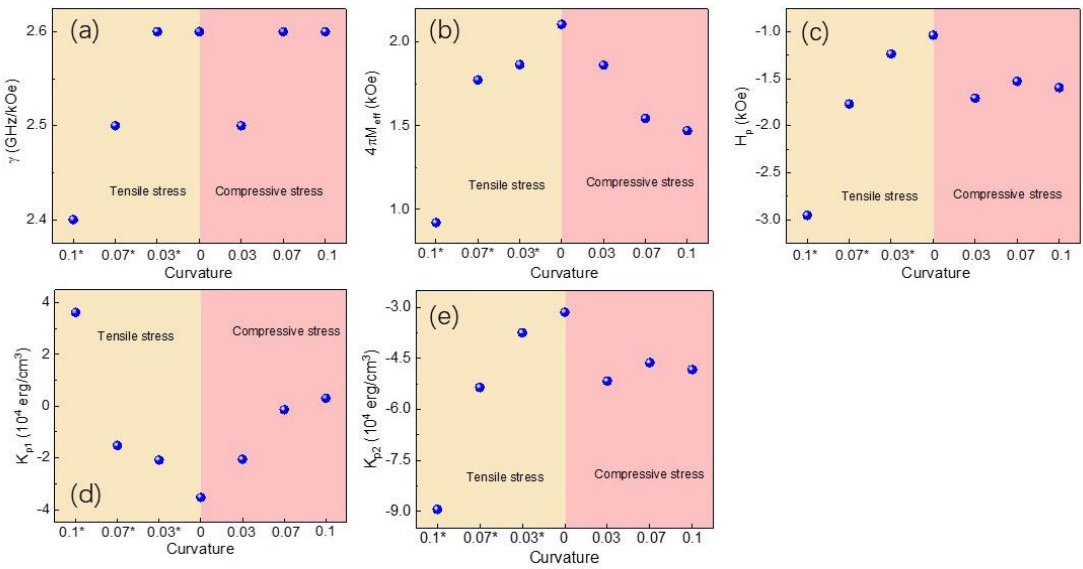

# Angular dependent ferromagnetic resonance of exfoliated yttrium iron garnet film under stress

Yufeng Wang, Peng Zhou[*], Shuai Liu, Yajun Qi[*], Tianjin Zhang[*]

Ministry of Education Key Laboratory for Green Preparation and Application of Functional Materials, Hubei Provincial Key Laboratory of Polymers, Collaborative Innovation Center for Advanced Organic Chemical Materials Co-constructed by the Province and Ministry, School of Materials Science and Engineering, Hubei University, Wuhan 430062, PR China

[*]E-mail addresses:

zhou@hubu.edu.cn (Peng Zhou),

yjqi@hubu.edu.cn (Yajun Qi)

zhangtj@hubu.edu.cn (Tianjin Zhang)

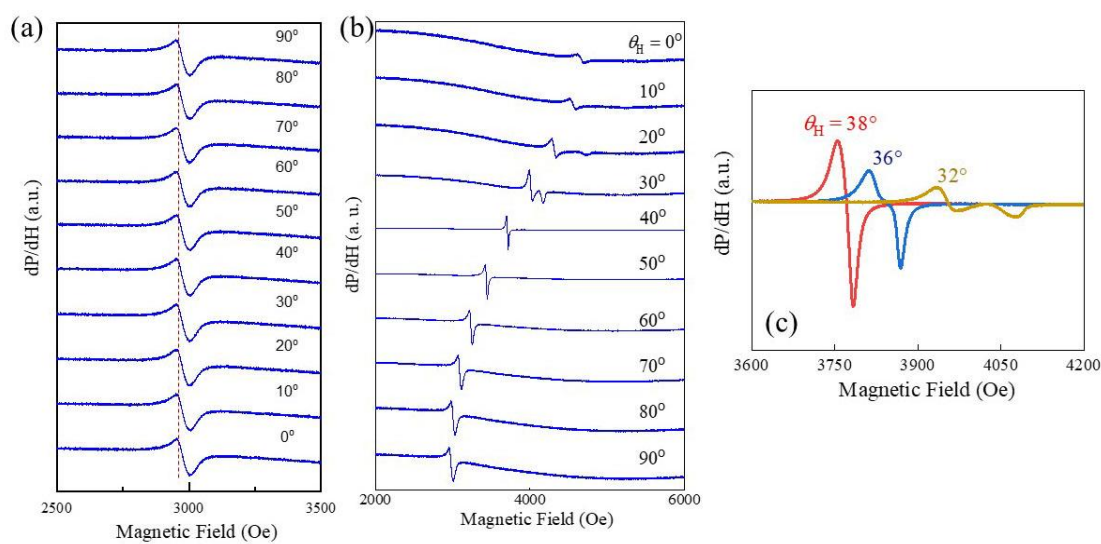

Figure S1 (a) In-plane and (b) out-of-plane angular dependent FMR spectra of YIG/2Gr/GGG, (c) FMR spectra of YIG/2Gr/GGG at $\theta_H$ = 32°, 36°, and 38°

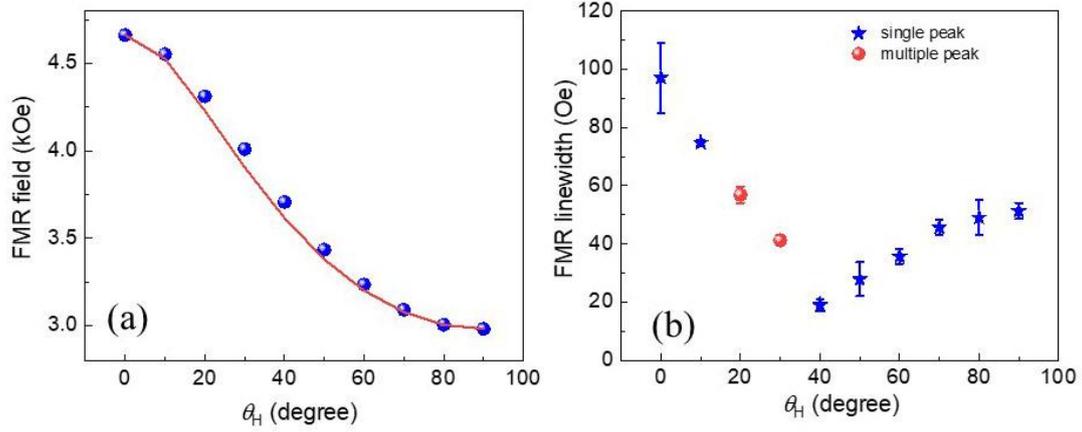

Figure S2 FMR field (a) and FMR linewidth (b) as a function of $\theta_H$. The red line in (a) is a fit to equation (1)

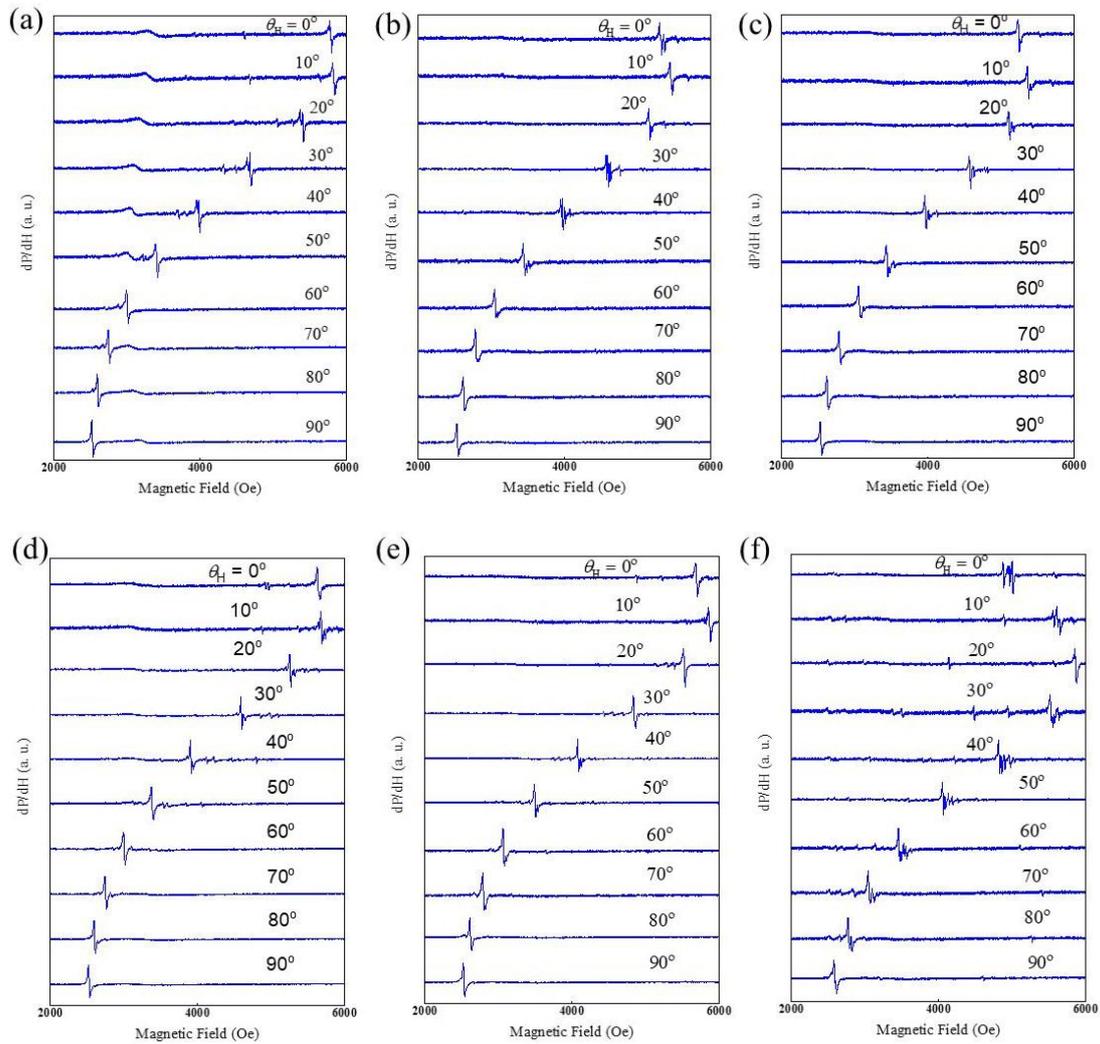

Figure S3 Angular dependence of FMR spectra of YIG/PET at curvature of (a) 0.03, (b) 0.07, (c) 0.1, (d) 0.03*, (e) 0.07*, (f) 0.1*